\documentstyle[11pt,aas2pp4]{article}

\newcommand{\etal}{{\rm et al.~}}            

\begin{document}

\title{Dust Properties of NGC~4753}
\author{G. C. Dewangan \altaffilmark{1}, K. P. Singh \altaffilmark{2}, P. N. Bhat \altaffilmark{3}}
\affil{
Tata Institute of Fundamental Research, Mumbai 400005, INDIA}
\altaffiltext{1}{gulab@tifr.res.in}
\altaffiltext{2}{singh@tifr.res.in}
\altaffiltext{3}{pnbhat@tifr.res.in}
\slugcomment{Version Date: May 5, 1999}

\begin{abstract}
We report $BVR$ surface photometry of a lenticular galaxy, NGC~4753 with prominent dust
lanes. We have used the multicolor broadband photometry to study dust-extinction as a function 
of wavelength and derived the extinction curve. We find the extinction curve of NGC~4753  
to be similar to the Galactic extinction curve in the 
visible region which implies that the sizes of dust grains responsible for optical 
extinction are similar to those in our Galaxy. We derive dust mass from optical 
extinction as well as from the far infrared fluxes observed with {\it IRAS \footnote[4]
{Infrared Astronomical Satellite}}. The ratio of the two dust masses, 
$\frac{M_{d,IRAS}}{M_{d,optical}}$, is 2.28 for NGC~4753, which is significantly lower 
than the value of  $8.4\pm 1.3$  
found previously for a large sample of elliptical galaxies. The total mass of the 
observed dust within NGC~4753 is about a factor of $10$ higher than the mass of dust
expected from loss of mass from red giant stars and destruction by sputtering and 
grain-grain collisions in low velocity shocks, and sputtering in supernova driven blast waves.  
We find evidence for the coexistence of dust and H$\alpha$ emitting gas 
within NGC~4753. The current star formation rate of NGC~4753, averaged over past
$2\times10^{6}~yr$, is estimated to be less than $0.21M_{\sun}yr^{-1}$. A substantial
amount of dust within NGC~4753 exists in the form of cirrus. 

\end{abstract}

\section{Introduction}
  It has now been well established that the presence of dust 
in early-type galaxies is the rule rather than an exception. 
Cool interstellar dust has been observed by IRAS (Jura~1986) and by means
of optical extinction (Hawarden \etal 1981; Sadler \& Gerhard~1985; 
Ebneter \etal 1988; Goudfrooij \etal 1994b). Using IRAS data, Knapp \etal (1989) have 
shown that about $45\%$ of the ellipticals and $68\%$ of the SO galaxies have been detected
at $60~\micron$ and $100~\micron$ bands while Goudfrooij \etal (1995) have reported
a detection rate of $61\%$ for their sample of Shapley-Ames elliptical
galaxies. The Hubble Space Telescope (HST) survey of ellipticals in the 
Virgo cluster has 
revealed the presence of dust in the nuclei of almost every galaxy (Jaffe \etal 1994).

Apart from dust, early-type galaxies also contain substantial amounts of gas at various
temperatures. Cold gas ($T\le 100~K$) has been observed via its HI $21~cm$ line emission
(Knapp \etal 1985) and CO line emission (Wiklind \& Rydbeck~1986; 
Phillips et.al~1986); warm gas ($T\approx 10^{4}K$) via optical emission 
lines (e. g. [OII], [OIII]) (Caldwell~1984; Phillips \etal 1986) and hot gas via their
X-ray emission (Forman \etal 1985; Canizares \etal 1987; Fabbiano \etal 1992).
The interstellar medium (ISM) of early-type galaxies, thus contains many different phases from cold to very hot.
Typically they contain an amount of cool dust of mass  $\simeq 
10^{4}- 10^{6}M_{\sun}$ (Jura \etal 1987; Goudfrooij \etal 1994c), hot 
gas of mass $\simeq 10^{9}-10^{10}M_{\sun}$ (Forman \& Jones~1985; 
 Fabian \etal 1987), and small amounts of warm gas of mass $\simeq  
 10^{3}-10^{5}M_{\sun}$  (Phillips \etal 1986;
 Caldwell~1984). 

The study of properties of different phases of ISM in 
early-type galaxies provides important clues to the origin, nature and fate of 
the interstellar matter. In particular, dusty early-type galaxies provide 
suitable environment to study the nature
of extra-galactic dust grains, either via their far infrared (FIR) emission or optical
extinction. The data can be used to study the physical mechanism operating on
them which can determine their size distribution, temperature, abundances, formation,
destruction etc. The
physical properties of the dust grains are functions of time and can be used
as an indicator for the time elapsed since the dust was last substantially
replenished (see Goudfrooij \etal 1994c).

 The emission line regions
in X-ray bright early-type galaxies are expected to be dust-free in view of very short life 
time ($\simeq 10^{6} - 10^{7}yr$) of dust grains. However, Goudfrooij \etal (1994c) 
found that the 
emission line regions in these galaxies are essentially always associated with
substantial dust-extinction. This dilemma can be resolved in the evaporative
flow scenario (de Jong \etal 1990) in which the observed dust immersed in hot, 
high pressure plasma may be replenished by the evaporation of cool clouds. The presence of cool 
clouds is assumed to be due to the capture of a companion 
dwarf galaxy, rich in dust and gas. A recent study of nuclear dust in 64 elliptical
galaxies imaged with HST has shown that the dust and gas are generally dynamically 
decoupled (van Dokkum \& Franx~1995). Moreover, the kinematics of
stars and gas in ellipticals with large dust lanes are found to be decoupled
(e.g. Bertola \etal 1988). These conclusions generally indicate an external origin
of dust in these galaxies. Further, the dynamical state of dust and gas
can be used to probe the intrinsic shape of the underlying galaxy. The dust
and gas, settled in the galaxy potential, are allowed in stable closed orbits 
which indicates a plane in the galaxy (Gunn~1979; Merrit \& de Zeeuw~1983; 
 Habe \& Ikewchi 1985, 1988).

In this paper, we present a detailed study of dust properties of an early-type galaxy
NGC~4753 with very prominent dust lanes. The galaxy NGC~4753 is located at a distance of $8.7~Mpc$ as determined from a distance modulus of $29.7$ by Buta \etal (1985) using published estimates and the light curve of supernova SN~1983g observed within the galaxy. NGC~4753 has been classified as a peculiar SO type 
galaxy in the Hubble atlas (Sandage~1961) because its underlying luminosity
distribution resembles an SO type galaxy. However, the third reference catalogue (RC3)
(de Vaucouleurs \etal 1991) classifies it as an irregular type, I0. The complex dust lanes 
in NGC~4753 pass through its center. Steiman-Cameron \etal (1992) have shown using $R$ 
band photometry that the dust lanes lie in a disk which is strongly 
twisted by differential precession.
We have carried out $BVR$ surface photometry of NGC~4753 and derived 
its color and extinction maps. Based on these, we 
estimate the dust mass from optical extinction. We have also 
derived the dust mass based on FIR emission.
The observed optical (RC3) and infrared (Knapp \etal 1989) properties of NGC~4753 are summarized 
in Table 1. The paper is organized as follows. In the next section, we provide 
details of our observation. In $\S 3$ we present our method of analysis and the 
results obtained. In $\S 4$ we discuss our results followed by conclusions in $\S 5$.

\section{Observations}
 NGC~4753 was observed with the Vainu Bappu Telescope (VBT) on the night of 
March 13, 1996.  The observations were carried with a liquid Nitrogen-cooled 
TEK $1024\times1024$ CCD chip placed at the prime
focus of the $2.3~m$ reflector. 
The nominal value of f-ratio is f/3.25. 
The pixel size of $24~\micron$ square per pixel of the CCD chip gives a scale 
of $0.6\arcsec$ per pixel and a total field of $10.24\arcmin\times 10.24\arcmin$. 
The observations were carried out under photometric conditions. The seeing ({\it FWHM}) 
was in the range $2.1\arcsec -  2.6\arcsec$. The exposure times were chosen not to saturate the CCD pixels 
because of the presence of bright stars in the neighbourhood of the galaxy NGC~4753. In 
order to achieve a good signal-to-noise ratio, two images of NGC~4753 were taken in each 
of the broad band filters B,V and R with exposure times 300, 180 and 60\,s  respectively. 
To correct for the bias level, bias images were taken just before and after the galaxy 
observation apart from many bias frames taken during the observing night. Several flat 
field images in each filter were taken by exposing to the twilight and dawn sky in 
order to correct for the non-uniformity of response of pixels of the CCD. The standard 
star field in the ``dipper asterism'' region of the open cluster M~67 was observed for 
photometric calibration. Both NGC~4753 and M~67 were observed close to the zenith.

\section{Analysis and Results}
   We have used IRAF\footnote[5]{IRAF
is distributed by the National Optical Astronomy Observatories, which is
operated by  the Association of Universities, Inc. (AURA) under cooperative
agreement with the National Science Foundation. The IRAF version $2.11.1$ was used.}
software package for the basic reduction and analysis of CCD images. Examination of the bias frames 
showed that the mean bias level did not vary significantly over the observation run. 
An average bias frame, constructed from bias frames taken very close to the 
galaxy observation, was subtracted from each of the object frames and flat field frames. 
The bias-subtracted frames were trimmed to a size of  $700\times850$ pixels in order to avoid 
the effects due to vignetting at the edges. The pixel-to-pixel response 
variation in each of the object frames were corrected by dividing the 
object frames by master flat frames in each band separately. The master 
flat frame in a filter band was obtained by averaging the best flat frames 
in the same filter and normalizing by the mean intensity level of the averaged frame. 
Cosmic ray events, seen as few isolated bright pixels, 
were removed by replacing them by the average intensity of four nearest neighbors. 
The two frames of NGC~4753, obtained in each band, were combined after alignment which 
increased the signal to noise ratio. The point spread functions of individual and 
combined frames in each band were found to be similar to an accuracy better than $0.5\%$ which imply that the alignments 
were correct. The pixel coordinates in each of the object image were converted into the 
standard equatorial coordinates for the epoch $2000$. The plate solutions were computed
by fitting a quadratic polynomial between the known celestial coordinates 
and pixels coordinates of stars nearby to
the object after projecting the celestial coordinates onto the plane tangent to the object center.
The subsequent analysis utilized the final corrected and combined object images. The 
bias-corrected and flat-fielded $B$ and $R$ band images of NGC~4753 are shown in 
Figures \ref{f1} and \ref{f2} respectively, where the effect of extinction due to dust can easily 
be seen as the severe departure of the isophotes from the nearly elliptical shapes 
at the positions of dust lanes or patches. It is also clear that the extinction 
due to dust is larger in the $B$ band as compared to that in the $R$ band.   

\subsection{Photometric Calibration}
 Many authors (e.g. Chevalier \& Ilovaisky~1991; Mayya~1991; Anupama \etal 1994; Bhat \etal 1992) have 
emphasized the advantages of using the standard star field in the ``dipper
 asterism'' region of the open cluster M~67 for photometric calibration. We have 
determined instrumental magnitude of stars in the ``dipper asterism'' region 
of M~67 by profile fitting photometry using ``DAOPHOT'' routine within IRAF. 
The method is discussed in detail by Stetson~(1987). To correct the instrumental 
magnitudes for atmospheric extinction, we have made use of extinction coefficients 
given by Mayya~(1991), who, based on eight nights observation at VBT
 during January, 1991 to April, 1991, has determined the average extinction coefficients 
for the observatory. The extinction corrected instrumental magnitudes and colors were 
transformed into the standard $BVR$ system by fitting the following equations to the data:
\begin{equation}
V-v_{0} = \alpha_{v}+\beta_{v}(B-V)
\end{equation}
\begin{equation}
B-V =  \alpha_{b-v} + \beta_{b-v}(b-v)_{0}
\end{equation}
\begin{equation}
V-R = \alpha_{v-r} + \beta_{v-r}(v-r)_{0}
\end{equation}
\begin{equation}
B-R = \alpha_{b-r} + \beta_{b-r}(b-r)_{0}
\end{equation}
where the upper case letter $B$, $V$ and $R$ are the standard magnitudes in the 
corresponding filters, taken from Chevalier \& Ilovaisky~(1991). The linear regression 
model chosen to fit the data was ordinary least squares regression of y on x (OLS(Y$\mid$X)), 
which has been recommended for the data used for calibration purposes (see Isobe \etal 1990; 
Feigelson \& Babu~1992). The derived transformation coefficients are given
 in Table 2.

\subsection{Surface Photometry}
 We have carried out isophotal analysis of the galaxy NGC~4753 in $B$, $V$ and $R$ bands. The shapes of the 
isophotes were analyzed using the ellipse-fitting routine within the 
``STSDAS'' \footnote[6]{The Space Telescope
Science Data Analysis System STSDAS is distributed by Space Telescope Science Institute. 
The STSDAS version 2.0.1 was used.} software package (for details, see Jedrzejewski~1987). 
A proper background subtraction is crucial in the analysis. We determined the median
intensity level (and its associated dispersion) in several $25\times25$ pixels boxes in the source free regions of an object image. The dispersion in intensity levels in each box was found to be less 
than $2.4\%$, $2.0\%$, and $2.1\%$ in $B$, $V$, and $R$ band images respectively, however, the dispersion in the median intensity levels was found to be $0.26\%$, $0.4\%$, and $0.19\%$ in $B$, $V$, and $R$ band images respectively.
We did not find any significant gradient in the median intensity levels in the source free regions across any of the object images. The sky background level was determined
as the mean of the median background intensity levels in each band separately.  
The background level, thus determined, 
was subtracted from
the object image in each band. 
The regions covered by stars and dust were masked and excluded from the isophotal analysis. The 
deviant pixels with intensity values $3\sigma $ below or above the mean intensity level 
were clipped and the maximum accepted fraction of flagged pixels 
in an ellipse fitting was kept at $50\%$.  The 
complex dust lanes and/or patches pass through the center of the galaxy, so it was 
not possible to determine the center very accurately. We choose the R band image, which 
is the least affected by dust-extinction, for detailed isophotal analysis. Starting with trial values of ellipticity ($\epsilon$), 
position angle ($\theta$) and ellipse center, an ellipse of mean intensity at a 
given length ($a$) of semi-major axis was fitted.  
The first two harmonics of the deviations 
from the trial ellipse were found. The best fitted ellipses were determined after 
performing a minimum of $10$ iterations and a maximum of $1000$ iterations to 
minimize the deviations. The third and fourth harmonics of the residual intensity from 
the best fitting ellipse were then evaluated. The procedure was repeated after changing 
the semi-major axis length by $10\%$, taking annuli and using the median value of the pixels 
for sampling along the elliptical path. The center was then determined by averaging the centers 
of the best fitted ellipses. The center, thus determined, was kept fixed and above described 
ellipse fitting procedure was repeated. The same center as determined for the $R$ band image 
was also used in the isophotal analysis of well aligned images in other bands.

The generated radial distribution of surface 
brightness and isophotal shape parameter profiles are shown in Fig. \ref{f3}. The position angle 
profile reveals twisted isophotes. The ellipticity profile shows that the isophotes become 
progressively more elliptical in the outer regions. The parameters $a3$, $a4$, and $b3$, $b4$ are the amplitudes of $sin3\theta$, $sin4\theta$, and $cos3\theta$, $cos4\theta$ coefficients, respectively,  
 of the isophotal deviation from the perfect ellipticity. 
It should be pointed out that in the inner 
regions ($a \leq 10\arcsec$ ) of NGC~4753, covered by complex dust lanes and/or 
patches, more than $50\%$ of the pixels are affected by dust-extinction. The isophotal 
parameter profiles in the inner region, therefore, should not be taken very seriously. 
It has been tried to estimate the correct dust-free intensity by decreasing the accepted 
fraction of pixels to fit ellipses in the inner regions. This may have affected the actual 
shape parameters in the central regions. The core region of the galaxy of size of order 
seeing disk was excluded from the isophotal analysis as these are affected by the seeing.

\subsection{Color Index maps}

We have generated color index maps ($B-R$, $\bv$, $\vr$) using the broad band images 
to find out the distribution of dust. Instrumental colors were corrected for atmospheric 
extinction using the average extinction coefficients of Mayya~(1991) and then converted 
into standard colors using the transformation coefficients given in Table 2. The $B-R$ color 
index map is shown in Fig. \ref{f4}, superimposed on which are the contours 
of $H\alpha$ image  (Singh \etal 1995). 
The brighter regions represent the part of the galaxy which are redder in color 
and hence represent cooler/dusty regions within the galaxy. The $B-R$ color in the dust-occupied 
regions within the galaxy was found to vary from $1.92$ to $2.68$. The maximum $B-R$ color 
is found to be $2.68$ at the center. In the dust-free regions  , the $B-R$ color as derived 
from the isophotal analysis was found to vary from $1.87\pm0.07$ at a semi major axis 
length of $6.0\arcsec$ to $1.38\pm0.07$ at the semi major axis length of $246.7\arcsec$. In the dust-free regions, the color gradient with respect to logarithmic galactocentric radius $r$, $\frac{d(B-R)}{d(logr)}$, was found to be $-0.21\pm0.05$ for NGC~4753. This value can be compared with the color gradients ($\frac{d(B-R)}{d(logr)}$) $-0.26$, $-0.23$, $-0.14$ for the SO galaxies NGC~3414, NGC~3607, NGC~5866 respectively as derived by Vader \etal 1988. Hence, the average color gradient of NGC~4753 in the dust-free regions   is normal for SO galaxies.
 Figure \ref{f4} also suggests the coexistence of dusty regions with the H$\alpha$-emitting gas. 

\subsection{Extinction Maps}

Several authors have studied dust properties of elliptical galaxies by an 
indirect method in which dust-extinction is determined as a function of 
wavelength by comparing the actual distribution of intensity of the galaxy 
with that expected in the absence of dust-extinction. 
The dust-free intensity distribution is modeled using elliptical isophotes. Sahu \etal (1998) have used 
the same method to study dust properties of an SO galaxy NGC~2076. The projections of three dimensional intensity profiles of bulge and disk of an SO galaxy onto the plane of sky are elliptical bulge and elliptical disk isophotes. If the disk is inclined with respect to the plane of the sky, disk isophotes are more elliptical than bulge isophotes. In many cases, the bulge and disk isophotal parameters are different. 
In the presence of a disk, the isophotes are not perfect ellipses. The deviations of isophotes of NGC~4753 from perfect ellipses are revealed by the presence of third and fourth order harmonics ($a3$, $b3$, $a3$, and $b4$).  Therefore, the projected intensity distribution of bulge and disk can be modeled by using the isophotal parameters and the harmonics. The dust-free model images of NGC~4753 were generated by interpolating the fitted isophotal parameters (dust-free) with polynomial of order three including the third and the fourth order harmonics determined above, thus incorporating the disk-like structure. In order to check, whether the dust-free intensity distribution of disk and bulge has been correctly modeled, we have carried out isophotal analysis of the dust-free model images of NGC~4753, and find that the profile of the parameter $b4$ derived from the isophotal analysis of dust-free ``model image" of NGC~4753 is very similar to the profile of the parameter $b4$ derived from the isophotal analysis of the ``observed image" of NGC~4753. Therefore, we conclude that the modeled dust-free intensity distribution of NGC~4753 has not been affected significantly by the presence of a stellar disk and the model image represents the actual dust-free intensity distribution of NGC~4753.  
The model images were used to create extinction maps in magnitude scale as follows:
\begin{equation}
A_{\lambda}=-2.5log_{10}(I_{\lambda,obs}/I_{\lambda,model})
\end{equation}
where $\lambda$= $B$, $V$, $R$. Note that absolute flux calibration is not needed for this purpose. The extinction map of NGC~4753 in $B$ band is shown in Fig. \ref{f5}. 
The brighter areas within the galaxy represent regions of large optical depth associated 
with the dust-extinction. The extinction in $B$ band was found to vary 
from $0.881\pm0.036$ to $0.438\pm0.058$. This estimate excludes central 
regions ($a \le 10\arcsec$) because of the poor accuracy of the ellipse-fitting procedure. 
Figure \ref{f5} also shows overlapped contours of $H\alpha$ image, which  demonstrates 
that H$\alpha$-emitting  regions and dust-occupied regions are co-spatial. The extinction 
maps in different bands were used to study the dust properties, as described in the next section.

\subsection{Extinction Curve}

 To quantify the extinction due to dust, we selected regions in the extinction 
map obviously occupied by dust.  
The numerical values of extinction, 
$A_{\lambda}$ $(\lambda$ = $B$, $V$, $R$), were calculated as the mean extinction within square 
boxes of size $5\times 5$ pixels, which is comparable to the size of the seeing disk. The uncertainties associated with the derived extinction values were estimated from the pixel-to-pixel scatter 
within each box.   
Since the isophotal model fits were mostly unreliable in the central 
regions ($a \le 10\arcsec$), we considered safer to exclude these regions 
from the estimation of extinction values. The average extinction values 
over the dust-occupied regions ($a \geq 10\arcsec$) within the galaxy image were assigned to the dust-occupied central regions. Since contribution from H$\alpha$ and [NII] line emission to the $R$ band are mostly concentrated in the central regions ($a \le 10\arcsec$) of NGC~4753, the derived extinction values in $R$ band are not significantly affected by the presence of line emission.

To determine the ratio of total extinction to selective extinction, 
we fitted BCES (Bivariate Correlated measurement Errors and intrinsic Scatter) least squares regression lines between different extinction 
values. The BCES method is a direct generalization of the ordinary least 
squares (OLS) regression method, modified to accommodate the measurement 
errors either correlated or uncorrelated and intrinsic scatter 
(for details, see Akritas \etal 1996). In the presence of errors, the 
use of OLS regression method can cause considerable bias, since it does not take 
into account measurement errors in the variables. Apart from the measurement 
errors in the derived extinction values, there are also intrinsic scatter 
depending on spatial distribution of dust and their properties. The measurement 
errors for extinction in two bands are uncorrelated. As a consequence, we 
must use the BCES technique ignoring the correlated errors in order to derive 
the linear regression lines relating extinction in two bands.
We fitted the BCES regression lines $A_{y}$ on $A_{x}$ and $A_{x}$ on $A_{y}$ ($x,y=B,V,R$; $x\ne y$).
The bisector of these two regression lines was chosen as the best fit line. 
The bisector treats the variables 
symmetrically and has been recommended for scientific problems where the goal is 
to estimate the underlying functional relationship between the 
variables (see Isobe \etal 1990; Feigelson \& Babu~1992). In order to compare our results
and to see the effects of errors on the regression coefficients, we have also 
fitted OLS regression lines using the method described by Isobe \etal (1990) between 
extinction values in different bands. The BCES and OLS regression coefficients 
are given in Table 3 and the fitted straight lines are shown in Figs. \ref{f6}(a) to (d). 
In Fig. \ref{f6}(a), the lines marked as 1, 2, and 3 are the linear regression fits
of $A_{B}$ on $A_{V}$, $A_{V}$ on $A_{B}$, and the bisector of the two lines  
respectively derived using the BCES technique. In Fig. \ref{f6}(b), the 
lines 1, 2, and 3 are same as those in Fig. \ref{f6}(a) but derived using the OLS method. As 
can be seen in Table 3, the lines marked 1 in Figs. \ref{f6}(a) and (b) are 
significantly different. It is also seen in Table 3 that the lines marked 2 in 
Figs. \ref{f6}(a) and (b) are significantly different but the BCES bisector and OLS bisector 
lines marked 3 in Figs. \ref{f6}(a) and (b) are similar.  
The difference in lines 1 in 
Figs. \ref{f6}(a) and (b), and also the difference in lines 2  can be explained as 
the effect of large errors as shown in Fig. \ref{f6}(a).
The BCES bisector and OLS bisector yielded intercepts close to zero which are 
within one standard deviations as expected. This further suggested the validity 
of use of bisector line between extinction values in different bands.  Henceforth, we 
use the BCES bisector line as the best fit regression line. 
The linear regression fit 
between $A_{R}$ and $A_{B}$  using the BCES bisector method is shown in Fig. \ref{f6}(c). Similarly, the BCES bisector line of $A_{V}$ and $A_{R}$ is shown in Fig. \ref{f6}(d).

The best fitting BCES bisector slopes and their associated uncertainties were subsequently used to derive the ratio of total extinction to 
selective extinction, $R_{\lambda} =\frac{A_{\lambda}}{A_{B}-A_{V}}$, and their associated 
uncertainties. The average extinction curve for the areas occupied by dust 
in NGC~4753 is shown in Fig. \ref{f7} along with the extinction curve of our Galaxy 
for comparison. The total extinction to selective extinction values for the Milky Way were 
derived in the same manner as for NGC~4753 from the ratios of extinction which, 
for the Galaxy, were taken from Rieke \& Lebofsky (1985). Figure \ref{f7} shows that 
the ratio $R_{\lambda} =\frac{A_{\lambda}}{A_{B}-A_{V}}$ varies linearly with 
inverse wavelength which is consistent with the result that for small grain 
size $x<1$, $Q_{ext}\propto \lambda^{-1}$, where $x=\frac{2\pi a}{\lambda}$, $a$ is the grain radius and $Q_{ext}$ is the extinction efficiency. From Fig. \ref{f7} it is obvious that the extinction curve for NGC~4753 is very similar to the Galactic extinction curve in the visible region. Also, the ratio of total extinction to selective
 extinction, $R_{V}$ = $3.1\pm0.3$ for NGC~4753,  is similar to 
the Galactic value of $3.1$. The extinction curves in Fig. \ref{f7} imply that the grain size
distribution within the dusty regions of NGC~4753 is similar to that of our Galaxy. This conclusion about the grain size distribution can be used to determine the total dust mass within NGC~4753, as discussed below.

\subsection{Dust Mass from Optical Extinction}

We have used the method of Goudfrooij \etal (1994c) to derive the dust mass from optical extinction values.
The total dust mass can be estimated by integrating the dust column density, given by
\begin{equation}
\Sigma_{d}=\int_{a_{min}}^{a_{max}} \frac{4}{3}\pi a^{3}\rho_{d} n(a) da \times l_{d}
\end{equation}
over the dust-occupied areas. In equation 6, $\rho_{d}$ is the grain mass 
density, $n(a)$ is the grain size distribution function, $a_{min}$ and $a_{max}$ 
are the lower and upper cutoffs of the grain size distribution. The dust 
column length, $l_{d}$, along the line of sight is determined from
\begin{equation}
A_{\lambda}=1.086C_{ext}(\lambda)\,\times l_{d}
\end{equation}
where $A_{\lambda}$ is extinction in magnitude at wavelength $\lambda$, and $C_{ext}(\lambda)$ is the extinction cross section of spherical grains per unit volume. $C_{ext}(\lambda)$ can be written as 
\begin{equation}
C_{ext}(\lambda)=\int_{a_{min}}^{a_{max}}Q_{ext}(a,\lambda)\pi a^{2}n(a)da
\end{equation}
where $Q_{ext}(a,\lambda)$ is the extinction efficiency  of grains of radii $a$ at wavelength $\lambda$. In computing $C_{ext}(\lambda)$, contribution of different chemical compositions of the dust grains is taken into account.

Since the extinction curve of NGC~4753 is quite similar to that of our Galaxy, we have employed the grain size distribution of Mathis \etal (1977), i.e.,
\begin{equation}
\nonumber
 n(a)da = A_{i}n_{H}a^{-3.5}da  ~~~~~~ (a_{min}\leq a\leq a_{max})
\end{equation}
with $a_{min}=0.005\micron$ and $a_{max}=0.22\micron$ in case $R_{V}=3.1$. In above 
equation,
$n_{H}$ is the number density of H nuclei, $A_{i}$ is the abundance of 
grains of type $i$ per H atom. As to the composition of grains, we assume 
spherical grains that are composed of either graphite or silicates, with equal 
abundances. This is justified by the fact that the Galactic extinction curve can be fitted assuming the above composition (Mathis \etal 1977). We have used parameterized $V$ band extinction efficiencies of Goudfrooij \etal 1994c for the 
two types of grains. The adopted values are 
\[ Q_{ext,silicate} = \left\{ \begin{array}{ll} 
              0.8\,a/a_{silicate} & \mbox{for $a < a_{silicate}$} \\
              0.8     & \mbox{for $a \geq a_{silicate}$}
              \end{array}
                \right.  \]
and
\[ Q_{ext,graphite} = \left\{ \begin{array}{ll}
              2.0\,a/a_{graphite} & \mbox{for $a < a_{graphite}$} \\
              0.8     & \mbox{for $a \geq a_{graphite.}$}
              \end{array}
                \right.  \]
with $a_{silicate}=0.1\micron$, and $a_{silicate}=0.05\micron$.
The values of various parameters used for estimation of total dust mass are given in Table 4. 
The average dust mass per pixel was computed using equations 6 to 9, 
and the average value of $A_{V}$.  
We have adopted a distance of $8.7\,Mpc$ for 
NGC~4753 (see Table 1). To find the total dust mass of NGC~4753, we integrated the average dust 
mass per pixel over all the dust-occupied pixels which were determined based on the 
criterion that a pixel within the galaxy image is considered to be dusty if the pixel value in the ratio of 
observed and smooth model image is $2\sigma$ or more below unity. The total mass of dust within NGC~4753, 
calculated thus, is $1.5\times 10^{5}\,M_{\sun}$. The mass of dust can also be determined 
based on far infrared (FIR) emission, as discussed in the next subsection.

\subsection{IRAS properties}

NGC~4753 was detected as a point source in the IRAS survey at all four bands $12\micron$, $25\micron$, $60\micron$ and $100\micron$. We obtained IRAS flux densities from Knapp \etal (1989). The flux densities at $60\,\micron$ and $100\,\micron$ were corrected for 
the contribution of circumstellar dust emission (Goudfrooij \etal 1995). The dust 
temperature was determined to be $30.4\,K$ based on the ratio of flux densities 
at $100\,\micron$ and $60\,\micron$ under the assumption that the FIR emission 
from the dust grains within NGC~4753 is governed by an emissivity law where emissivity
is proportional to $\lambda^{-1}$ at wavelengths $\lesssim$ $200\micron$ (Schwartz~1982; Hildebrand~1983; Kwan \& Xie~1992). Since a distribution of temperature may be more appropriate for dust, the derived dust temperature should be regarded as a representative value. We have estimated the dust mass of NGC~4753 following the method outlined by Hildebrand~(1983), Thronson \etal (1986) and Goudfrooij \etal (1995) which is based on measurements of FIR flux density $f(\nu)$, dust temperature $T_{d}$, dust emissivity $Q(\nu)$ and size of the grains. The total dust mass is given by
\begin{equation}
M_{d}=\frac{D^{2}f(\nu)}{B(\nu,T_{d})}.\frac{\frac{4}{3}a}{Q(\nu)}\rho_{d}
\end{equation}
where $\rho_{d}$ is the specific grain mass density, $a$ is the average grain radius weighted by the grain volume and D is the distance of the galaxy.
 The derived cool dust mass is $3.46\times 10^{5}\,M_{\sun}$. Assuming 
that the infrared spectrum of NGC~4753 can be fitted by a function of the form $f(\nu) \propto \nu B(\nu,T_{d})$ which is a good
approximation to the total infrared spectrum of most galaxies (Telesco \& Harper 1980; Rickard \& Harvey 1984), the total infrared luminosity can be written as (Thronson \etal 1986)
\begin{small}
\begin{displaymath}
L_{IR}(L_{\sun})=3.7\times 10^{-12}D^{2}f(\nu)T_{d}^{5}\lambda^{4}[exp(1.44\times10^{4}/\lambda T)-1]
\end{displaymath}
\end{small}
where $D$ is the object distance in $Mpc$, the flux density $f(\nu)$ is in $Jy$, and $\lambda$ is in $\mu m$. Using the above relation the total infrared luminosity of NGC~4753
is estimated to be $6.6\times 10^{8}\,L_{\sun}$.    

\section{Discussion}
The surface brightness profile and isophotal shape 
parameters (Fig. \ref{f3}) of NGC~4753 reveal some intensity features of this early-type galaxy. The parameter $b4$, the amplitude of $cos4\theta$
coefficient of the isophotal deviation from the best fitting ellipse, is
interpreted as being due to an embedded disk ($b4>0$) or due to boxiness
of isophotes ($b4<0$) (e.g. Peletier \etal 1990). The positive values of
$b4$ indicate a disk component in NGC~4753. The galaxy is very 
flattened in its outer parts. 
The presence of disk component is supported by large rotational velocities,
$v \ge 250 km~s^{-1}$ measured by Chromey~(1973). The disk component and the twisted isophotes as revealed by the position angle profile, Fig. \ref{f3}(b), are supported by a model 
where an inclined disk twisted by differential precession fits the observed dust
lanes lying on the disk very well (Steiman-Cameron \etal 1992).

We have derived the dust mass of NGC~4753 from optical extinction as well as from IRAS flux densities. The ratio of the two dust masses, $\frac{M_{d,IRAS}}{M_{d,optical}}$ is 2.28. The discrepancy can be attributed to
 the fact that the dust mass derived from optical extinction is a lower limit
because the extinction is observed only when the dust is in front of the stars. The
dust is probably distributed between the stars or stellar system, the actual
dust mass is expected to be higher than the dust mass derived from optical 
extinction. Goudfrooij \etal (1995) have reported that the average ratio
$\langle \frac{M_{d,IRAS}}{M_{d,optical}} \rangle$ to be $8.4\pm 1.3$ for a
large sample of elliptical galaxies for which the presence of dust is revealed by 
both FIR emission and optical dust lanes or patches. This ratio is significantly
higher than the ratio for NGC~4753 which implies that the 
dust mass discrepancy in NGC~4753
is not as severe as for an average elliptical galaxy. Sahu \etal (1998) have 
reported
that $\frac{M_{d,IRAS}}{M_{d,optical}}$ = 1.14 for an SO galaxy NGC~2076. This 
indicates that the difference in dust mass derived from FIR emission and 
optical extinction is not significant for SO galaxies in contrast to the dust
mass discrepancy in elliptical galaxies. This is an expected result because the
dust masses derived for spiral galaxies from optical and IRAS data reveal that
$M_{d,IRAS} \ll M_{d,optical}$ (Goudfrooij~1996)  and the SO galaxies, being intermediate in the morphological sequence from ellipticals to spirals, can be expected to have intermediate properties. 

A rough estimate of molecular gas content of NGC~4753 can be made from the dust mass derived from FIR flux if we assume that the ratio of molecular gas to dust mass to be in the range $50 - 700$ as given by Wiklind \etal (1995) for emission-line regions  . This would render the molecular gas mass of 
NGC~4753 to be in the range $0.2- 2.4\times 10^{8}M_{\sun}$. The masses of different contents of NGC~4753 are given in Table 5. 

The origin of the observed interstellar dust can be either internal (e.g. due to mass loss
from stars) or external (galaxy-galaxy interaction or merger). In the solar neighborhood,
the potential sources of dust grains are supernovae, red giant stars, protostellar 
nebulae and planetary nebulae with relative weights of $100$, $37$, $<16$, and $6$ respectively (Dwek \& Scalo 1980). In the case of early-type galaxies, the 
rate of star formation and hence the rate of supernova explosions are much lower than
the corresponding rates in spiral galaxies. The rate of supernova explosions in early-type galaxies is about a factor of $16$ smaller then the rate in spiral galaxies (Sbc-Sd) (Cappellaro \etal 1993). Therefore, it can be expected that the dominant sources of
dust grains in early-type galaxies are the red giant stars. In the following, we consider 
dust mass within NGC~4753 arising from red giant stars and neglect the other 
possible sources of dust grains. The
total mass loss rate from red giant stars within NGC~4753 can be estimated from the excess infrared (IR) luminosity arising from circumstellar shells of the red giant stars (see Soifer \etal 1986). The excess IR luminosity arising from circumstellar shells of red giant stars can be determined by comparing the IR luminosities from photospheres of normal stars and red giant stars and by counting the total number of red giant stars within NGC~4753. Since individual stars within NGC~4753 could not be resolved, it is not possible to determine the excess IR luminosity arising from circumstellar shells of the red giant stars. Hence the total mass loss rate from red giant stars within NGC~4753 cannot be determined directly. However, following the analysis of Jura \etal (1987) and assuming that stellar population within NGC~4753 is similar to that of elliptical galaxies or bulge of M~31, the total mass loss rate from red giant stars within NGC~4753 can be estimated. The total mass loss rate from red giant stars within an elliptical galaxy can be written as (Jura \etal 1987)
\begin{equation}
\dot{M}~(M_{\sun}~yr^{-1}) = 4.3 \times 10^{-30}~L_{\nu} (12\micron)~(erg~s^{-1} Hz^{-1})
\end{equation}
where $L_{\nu}(12\micron)$ is the luminosity per unit frequency at $12\micron$. For NGC~4753, we calculate $L_{\nu} (12\micron)$ to be $3.0\times 10^{26}~erg~s^{-1}~Hz^{-1}$ from the measured flux at $12\micron$ and adopting a distance of $8.7~Mpc$ (see Table 1). Substituting $L_{\nu} (12\micron) $ into above equation, we estimate the total mass loss rate from red giant stars within NGC~4753 to be $1.3\times 10^{-3}M_{\sun}~yr^{-1}$.
 The derived mass loss rate is uncertain because stellar population within elliptical galaxies and NGC~4753 may be significantly different, however, the mass loss rate is not expected to vary by large factors. If we assume that the gas-to-dust ratio in the circumstellar shells of red giant stars to be $100:1$ by mass, with a factor of $2-3$ uncertainty (see Knapp \etal 1993),
 the total loss rate of dust mass by red giant stars within NGC~4753 is about $1.3\times 10^{-5}M_{\sun}~yr^{-1}$. The  mass of dust accumulated within NGC~4753 can be estimated by the mass loss from stars and by the mechanisms of destruction of dust (see below). The rate at which dust mass is accumulated within NGC~4753 can be written as
\begin{equation}
\frac{\partial M_{d}(t)}{\partial t}=\frac{\partial M_{d,s}}{\partial t} - M_{d}(t) \tau^{-1}
\end{equation}
where $\frac{\partial M_{d,s}}{\partial t}$ is the loss rate of mass in the form of dust by red giant stars and $\tau^{-1}$ is the destruction rate of dust.
 The second term in the right represents the rate at which mass of dust 
within NGC~4753 is destroyed. Assuming that the mass loss from red 
giant stars begins at an age $\sim 10^{6}~yr$ of NGC~4753 
and taking $ \tau^{-1}\sim 7.14 \times 10^{-10}~yr^{-1}$ (see below), we have 
solved numerically the above equation for the mass of dust accumulated 
as a function of time. The build up of mass of dust within NGC~4753 is 
shown in Fig. \ref{f8}. Assuming an age of $10^{10}~yr$ for NGC~4753, the total 
mass of dust accumulated within NGC~4753 is calculated to be $1.8 \times 10^{4}M_{\sun}$,
which is about a factor of $10$ lower than the measured mass of dust. This indicates that the mass loss rate from red giant stars is probably not sufficient to account for the observed dust within NGC~4753. Based on our simplified calculation given above, the possibility of an external origin of dust within NGC~4753 can not be ruled out.

X-ray emission of varying amounts is seen in several early-type galaxies. 
Integrated X-ray emission from discrete sources and the presence of 
hot gas ($ \sim10^{7}K$), both can contribute to the overall X-ray emission. 
Hot plasma, if present, can destroy the dust very efficiently (Draine \& Salpeter 1979a). The ratio of total X-ray luminosity to absolute blue luminosity is a good indicator of the presence of hot gas in an early-type galaxy (Canizares \etal 1987).
Therefore, in order to estimate the dust destruction time scale, we examine X-ray 
emission from NGC~4753. The quantity $log(\frac{L_{X}}{L_{B}})$ ($L_{X}$ in $ergs~s^{-1}$ and $L_{B}$ in $L_{\sun}$) for NGC~4753 is 29.42, based on which Canizares \etal (1987) would conclude that the total X-ray luminosity of NGC~4753 is consistent with that expected from the integrated X-ray luminosity of discrete sources. The galaxy NGC~4753 is a member of the lowest $\frac{L_{X}}{L_{B}}$ group in Kim \etal (1992) where it has been shown that early-type galaxies in the lowest $\frac{L_{X}}{L_{B}}$ group have a very soft X-ray excess which amounts to about half the total X-ray emission. 
This emission could be due either to a cooler ISM or to the integrated emission of 
soft stellar sources. In any case, ISM of these galaxies is not heated to high X-ray 
temperatures. This implies that the dust in NGC~4753 is unlikely to be embedded 
in a very hot plasma. In the absence of hot plasma, the most effective destruction mechanism for
refractive grains (e.g., graphite and silicate) are sputtering and grain-grain collisions in 
low velocity shocks ($v_{shock}\le 60~km~s^{-1}$), and sputtering in supernova driven blast
waves (Goudfrooij \etal 1994c). Using the model of Draine \& Salpeter (1979b), appropriate for X-ray
faint early-type galaxies, and recent estimate for the rate of supernova explosions in early-type galaxies,
Goudfrooij \etal (1994c) have estimated the life time $\tau \sim 1.4\times10^{9}~yr$ for $0.1~\micron$
refractory grains. We have used this value to calculate the mass of dust
accumulated within NGC~4753, given above.

H$\alpha$+[NII] emission line regions in the center of NGC~4753 are embedded in the regions with substantial optical extinction (see Figs. \ref{f4} and \ref{f5}).This association of emission line regions and regions with substantial dust absorption points towards the coexistence of dust and warm gas ($\sim 10^{4}K$) within NGC~4753. The observed H$\alpha$ line emission can be accounted by photo-ionization from post asymptotic giant branch stars (see Singh \etal 1994; 1995), which can also supply significant amounts of dust.

As to the fate of cool dust and gas, we examine the possibility of star formation within NGC~4753. The galaxy occupies an intermediate position but close to the lower right end in the phenomenological IRAS color-color diagram of Helou~(1986), where both the cirrus component and active star forming regions contribute to the total FIR emission.
However, the FIR emission from active star
forming regions is less than $50\%$ of the total FIR luminosity, so that the total FIR emission of NGC~4753 can not be interpreted as due to current star formation. Following Thronson \& Telesco~(1980), the current star formation rate averaged over past $2\times10^{6}~yr$ can be calculated using the relation
\begin{equation}
\dot{M}_{FIR}=6.5\times10^{-10}L_{FIR}(L_{\sun})
\end{equation}
where $L_{FIR}$ is the total FIR luminosity. However,
 considering the phenomenological model of Helou~(1986), only the FIR luminosity due to the active star forming regions must be used in above expression. We estimate the current star formation
rate of NGC~4753, averaged over past $2\times10^{6}~yr$ to be less than $0.21M_{\sun}yr^{-1}$. As discussed above, the cirrus component of dust 
contributes more than $50\%$ to the total FIR luminosity, hence significant 
amount of dust within NGC~4753 is in the form of cirrus. This suggests that 
the cirrus clouds are not destroyed efficiently because of the lack of hot ISM.
Also, these regions are 
not sufficiently dense to lead to gravitational instability to form stars.  
\section{ Conclusions}

Our isophotal analysis of NGC~4753 has revealed a twisted disk component in it. 
The dust grain properties of NGC~4753 appear to be similar to that of our Galaxy based on comparison of the extinction curves.
The derived mass of cold dust in NGC~4753 is $1.5\times 10^{5}M_{\sun}$ of NGC~4753 
based on optical extinction, and $3.46\times 10^{5}M_{\sun}$ based on FIR fluxes. 
The discrepancy between the two dust masses for NGC~4753 is not as severe as the 
discrepancy generally found for the elliptical galaxies. The accumulated mass of dust within NGC~4753
from mass loss by red giant stars after taking into account the efficient destruction processes is $\sim 1.8\times 10^{4}M_{\sun}$, which is about a factor of $10$ lower than the measured dust mass. 
Dust and ionized gas coexist within NGC~4753. The current star formation rate of NGC~4753, averaged over past $2\times10^{6}~yr$, is derived to be less than $0.21~M_{\sun}yr^{-1}$ using the FIR emission arising from active star forming regions. A significant amount of dust within NGC~4753 seems to exist in the form of cirrus. 
\section{
Acknowledgments}

We thank the Director of the Indian Institute of
Astrophysics and the Time Allocation Committee for allotting the dark
nights for our observations.
The members of the technical support staff of the VBT are gratefully acknowledged for
their assistance during the observations. We thank an anonymous referee whose critical comments and suggestions helped in improving the paper. We also thank A. D. Karnik and D. K. Ojha for helpful discussions during the analysis.
This research has made use of the NASA/IPAC Extra-galactic Database (NED)
which is operated by the Jet Propulsion Laboratory, Caltech, under
contract with the National Aeronautics and Space Administration.

\newpage
\begin{figure}
\caption
{ Bias-subtracted, flat-fielded and cosmic ray removed $B$ band
image of NGC~4753 superimposed on which are the contours of same image. The contour levels are drawn at $4.2\%$, $4.5\%$, $5\%$, $6\%$, $7\%$, and $8\%$ of the peak intensity. The outermost contour is drawn at a surface brighness $23.1 ~magnitude ~per ~arcsec^{2}$. The brighter lanes or patches represent dust-occupied regions. }
\label{f1}
\end{figure}

\begin{figure}
\caption
{ $R$ band image of NGC~4753 overlapped with the contours of the same image. The contour levels are drawn at $3\%$, $3.3\%$, $4\%$, $5\%$, $6\%$, $7\%$ and $8\%$ of the peak intensity. The outermost contour corresponds to the surface brightness $22.1 ~magnitude ~per ~arcsec^{2}$.}
\label{f2}
\end{figure}

\begin{figure}
\caption
{ Results of $R$ band isophotal shape analysis of NGC~4753.
(a) Surface brightness profile, (b) position angle
profile, (c) ellipticity profile, (d) amplitude, $a3$, of residual
$sin3\theta$ coefficient of the isophotal deviation from perfect ellipse as a function of semi-major axis length (SMA). (e) amplitude, $a4$, of residual $sin4\theta$ coefficient of the isophotal deviation from perfect ellipse as a function of semi-major axis length (SMA). (f) amplitude, $b3$, of residual $cos3\theta$ coefficient of the isophotal deviation from perfect ellipse as a function f semi-major axis length (SMA). (g) amplitude, $b4$, of residual $cos4\theta$ coefficient of the isophotal deviation from perfect ellipse as a function of semi-major axis length (SMA). }
\label{f3}
\end{figure}

\begin{figure}
\caption
{ $B-R$ color image of NGC~4753 superimposed on which are the contours
of $H\alpha$ image.  The brighter regions represent part
of the galaxy which are redder in color. 
The maximum $B-R$ color is found to be 2.677 at the center. The H$\alpha$+[NII] contour are drawn at levels $5\%$, $7.5\%$, $10\%$, $15\%$, $25\%$, $50\%$ and $80\%$ of the peak intensity. The H$\alpha$+[NII] image is taken from Singh \etal (1995).
  }
\label{f4}
\end{figure}

\begin{figure}
\caption
{ Extinction image of NGC~4753 in $B$ band. The brighter areas represent
regions of large optical depth associated with the dust-extinction. The 
superimposed contours of H$\alpha$+[NII] (Singh \etal 1995) map are 
drawn at levels $5\%$, $7.5\%$, $10\%$, $15\%$, $25\%$, $50\%$ and $80\%$ 
of the peak intensity and show spatial association of dust and H$\alpha$ gas. 
}
\label{f5}
\end{figure}

\begin{figure}
\caption
{ BCES and OLS regression lines between extinction values in different pass bands. 
(a) BCES regression lines between extinction in $V$ and $B$ bands. The line 1 is the BCES regression line of $A_{V}$ on $A_{B}$, the line 2 is 
the BCES regression line of $A_{B}$ on $A_{V}$, and the line 3 represents 
the BCES bisector of the above two lines 1 and 2. (b) OLS regression lines between $A_{V}$ on $A_{B}$. The line 1 is the OLS line of $A_{V}$ on $A_{B}$, line 2 is the OLS line of $A_{B}$ on $A_{V}$, and line 3 is the OLS bisector of lines 1 and 2. (c) BCES bisector regression line of $A_{R}$ and $A_{V}$.
(c) BCES bisector regression line of $A_{R}$ and $A_{B}$. The quantities $A_{\lambda}$ ($\lambda$ = $B$, $V$, and $R$) represent the dust-extinction in the band $\lambda$.  }
\label{f6}
\end{figure}

\begin{figure}
\caption
{ Extinction curve of NGC~4753. The Galactic extinction curve is also plotted for comparison.
The quantity $R_{\lambda}$ ($ = \frac{A_{\lambda}}{A_{B}-A_{V}}$) is the ratio of total extinction to selective extinction in the band $\lambda$ = $B$, $V$, $R$. }
\label{f7}
\end{figure}

\begin{figure}
\caption
 {The total dust mass accumulated within NGC~4753 as a function of time.}
\label{f8}
\end{figure}

\end{document}